\documentclass[aps, twocolumn, showpacs, preprintnumbers, superscriptaddress, amsmath, amssymb, prb]{revtex4}
\usepackage{graphicx}
\usepackage{dcolumn}
\usepackage{bm}
\usepackage{color}
\usepackage{bbm}%

\makeatletter
\newsavebox\myboxA
\newsavebox\myboxB
\newlength\mylenA

\newcommand*\xoverline[2][0.75]{%
    \sbox{\myboxA}{$\m@th#2$}%
    \setbox\myboxB\null
    \ht\myboxB=\ht\myboxA%
    \dp\myboxB=\dp\myboxA%
    \wd\myboxB=#1\wd\myboxA
    \sbox\myboxB{$\m@th\overline{\copy\myboxB}$}
    \setlength\mylenA{\the\wd\myboxA}
    \addtolength\mylenA{-\the\wd\myboxB}%
    \ifdim\wd\myboxB<\wd\myboxA%
       \rlap{\hskip 0.5\mylenA\usebox\myboxB}{\usebox\myboxA}%
    \else
        \hskip -0.5\mylenA\rlap{\usebox\myboxA}{\hskip 0.5\mylenA\usebox\myboxB}%
    \fi}

\begin{document}

\title{Controllable Precision of the Projective Truncation Approximation for Green's Functions}

\author{Peng Fan}
\affiliation{Department of Physics, Renmin University of China, 100872 Beijing, China}

\author{Ning-Hua Tong}
\email{nhtong@ruc.edu.cn}
\affiliation{Department of Physics, Renmin University of China, 100872 Beijing, China}
\date{\today}

\begin{abstract}
Recently, we developed the projective truncation approximation for the equation of motion of two-time Green's functions (P. Fan et al., Phys. Rev. B {\bf 97}, 165140 (2018)). In that approximation, the precision of results depends on the selection of operator basis. Here, for three successively larger operator bases, we calculate the local static averages and the impurity density of states of the single-band Anderson impurity model. The results converge systematically towards those of numerical renormalization group as the basis size is enlarged. We also propose a quantitative gauge of the truncation error within this method and demonstrate its usefulness using the Hubbard-I basis. We thus confirm that the projective truncation approximation is a method of controllable precision for quantum many-body systems.
\end{abstract}



\maketitle

\section{Introduction}

Green's function (GF) is a widely used tool in the study of quantum many-body physics. Among many methods for calculating GF, the equation of motion (EOM) approach to the two-time GF is based on the Heisenberg equation of motion of operators\cite{Martin1,Bogolyubov1,Tyablikov1,Zubarev1}. For a given interacting Hamiltonian, the EOM of a given GF contains higher order GFs and repeatedly applying the EOM generates a chain of GFs. In the conventional Tyablikov-type truncation approximation\cite{Tyablikov1}, the GF of certain order is approximated as a linear combination of the lower order GFs in the frequency domain. This leads to a set of closed but approximate algebraic equations for the GFs, which can be solved to obtain the desired GFs. 

Decades of experience on this practice for various Hamiltonians shows that naive truncation of the EOM has some drawbacks. First, causality of the GFs is not guaranteed. The truncation may destroy the correct analytical structure of GF, i.e., GF containing only real simple poles. Second, for a given higher order GF, the truncation scheme is not unique. Different truncations may lead to drastically different results. Since there is no transparent clue for the optimal truncation scheme, in practice, the truncation depends heavily on experience. Due to these drawbacks, the EOM truncation approach is usually regarded as an uncontrolled approximation. Its application in the modern study of quantum many-body physics is therefore severely limited.

There are efforts to overcome the drawbacks of the EOM truncation approach. Using the idea of operator projection, Mori\cite{Mori1,Mori2} and Zwanzig\cite{Zwanzig1} developed the generalized Langevin equation formalism for the operator EOM, which can be used to calculate GFs. In particular, an elegant and exact continued fraction formalism was proposed to express GF or time correlation functions in terms of projecting coefficients in the operator space\cite{Mori2,Lee1,Lee2}. Similar theories have been developed by Tserkovnikov\cite{Tserkovnikov1,Tserkovnikov2} and applied by many other researchers under different names, including the two pole approximation\cite{Roth1}, composite operator approach\cite{Avella1,Avella2}, projection operator approach\cite{Fulde1,Fulde2,Imada1,Imada2}, irreducible GF method\cite{Kuzemsky1}, and many others\cite{Rowe1,Plakida1}. These closely related theories employ the operator projection idea to truncate the EOM. They have the advantage that the causality of GF is guaranteed by the formalism. Also, the time translation invariance of the equilibrium state is strictly obeyed by the GF. This is embodied by the fact that $\partial G(t-t^{\prime})/\partial t$ and $\partial G(t-t^{\prime})/\partial t^{\prime}$ give equivalent formula. Therefore, the first drawback of the EOM approach is removed.

However, in these theories, except for special cases\cite{Lee3}, the projection coefficients cannot be calculated without introducing additional approximations. The second drawback of the EOM truncation approach, i.e., the arbitrariness in the truncation, is still present in these theories. 
In our recent work~\cite{Fan1}, we proposed a practical and systematic projective truncation approximation (PTA) for the EOM of GFs. For a selected set of operator basis $\left\{ A_1, A_2, ..., A_n \right\}$, our theory is equivalent to the matrix form of the Mori-Zwanzig formula for GFs, with the memory function matrix neglected. By using a partial projection approximation, we reduce the calculation of projecting coefficients into that of two matrices, the inner product matrix ${\bf I}$ and the natural closure matrix ${\bf M}$. They are calculated self-consistently by the fluctuation-dissipation theorem (for ${\bf I}$) and through the commutators $[A_{i}, H]$ ($i=1,2,...,n$) (for ${\bf M}$), respectively. The arbitrariness in the truncation is thus removed. For the Anderson impurity model, our PTA at the same truncation level is superior to the conventional Lacroix approximation~\cite{Lacroix1} and the results are in quantitative agreement with those of the numerical renormalization group (NRG)\cite{Wilson1}.

In this paper, we study the convergence properties of PTA, by comparing the results of PTA on three successively larger bases. Using the Anderson impurity model and NRG results as reference, we examine whether the PTA results are improved with enlarging basis size and converge towards the exact ones. The positive results of this convergence check establishes PTA as a method of controlled precision for quantum many-body systems.

\section{Projective Truncation Approximation for EOM}

The projective truncation approximation for the EOM of GFs was developed in Ref.~\onlinecite{Fan1}. In this section, for the sake of completeness, we overview the general formalism.

For a given Hamiltonian $H$, we choose $n$ linearly independent operators $ \left\{ A_{1}, A_{2}, ..., A_{n} \right\}$ to span a subspace of the full Liouville space. To truncate the higher order operators generated by EOM, we project them into this subspace and neglect the component orthogonal to it. Therefore, the basis set should contain the most important excitations of the system. Such projective truncation becomes exact if the basis is complete.

For the column vector $\vec{A}$ formed by basis operators, the retarded GF matrix is defined as
\begin{equation}      \label{1}
 {\bf G} \left( \vec{A}(t) | \vec{A}^{\dag} (t^{\prime}) \right) = -\frac{i}{\hbar} \theta(t-t^{\prime}) \left\langle \left\{ \vec{A}(t),\vec{A}^{\dag}(t^{\prime})  \right\}\right\rangle
\end{equation}
where $\theta(t-t^{\prime})$ is the Heaviside step function and $\vec{A}(t)$ is the vector of basis operators in Heisenberg picture. In this paper we only consider the Fermion-type GF and the curly bracket in the above equation denotes anti-commutator. Below we take the natural unit and drop $\hbar$.

The equation of motion for the GF matrix in the frequency domain reads
\begin{eqnarray}      \label{2&3}
\omega G\left( \vec{A} | \vec{A}^{\dag} \right)_{\omega} &=& \langle \{ \vec{A}, \vec{A}^{\dag} \} \rangle + G\left( [ \vec{A}, H ] |\vec{A}^{\dag} \right)_{\omega} ,   \\
\omega G\left( \vec{A}|\vec{A}^{\dag} \right)_{\omega} &=& \langle \{ \vec{A},\vec{A}^{\dag} \} \rangle - G\left( \vec{A}|[\vec{A}^{\dag}, H ] \right)_{\omega}.   
\end{eqnarray}
We define the inner product of operators $X$ and $Y$ as
\begin{equation}      \label{4}
   (X|Y) \equiv \langle \{ X^{\dagger}, Y \} \rangle,
\end{equation}
where $\langle  \hat{O}\rangle = Tr(\rho\hat{O})$ and $\rho = e^{-\beta H} / Tr(e^{-\beta H} )$ is the equilibrium density operator of $H$ at temperature $T$. Eq.(4) fulfils the standard requirements for the inner product in a linear space. Other definitions of inner product can be found in the literature as well.~\cite{Mori1,Mori2,Rowe1}

We write the commutator between the basis operators and the Hamiltonian as
\begin{equation}    \label{5}
   [ A_{i}, H ] = \sum_{j} { \bf M} _{ji} A_{j} + B_i .
\end{equation}
The first term on the right-hand side includes all the basis operators that naturally appear in the commutator. ${\bf M}$ is called the natural closure matrix. $B_{i}$ is the newly generated operator outside the basis (In certain situations, $B_i$ may include basis operators for symmetry reasons, see below.). 
We further decompose
\begin{equation}      \label{6}
B_i = \sum_{j} { \bf N} _{ji} A_{j} + \delta B_{i} ,
\end{equation}
where $\delta B_{i}$ is the component orthogonal to the subspace of basis operators. That is, $\left(A_{k}| \delta B_{i} \right) =0$ for $i,k=1,2,..., n$. ${\bf N}$ is obtained by projecting Eq.(6) onto the basis operators and solving the obtained equation
\begin{eqnarray}    \label{7}
   {\bf P} = {\bf I N}, 
\end{eqnarray}
with ${\bf P}$ and ${\bf I}$ defined as ${\bf P}_{ij} \equiv (A_{i}| B_j)$ and ${\bf I}_{ij} \equiv (A_{i} | A_{j})$, respectively.

Putting Eq.(6) into Eq.(5) and projecting it onto the basis operators, we obtain
\begin{equation}      \label{8}
    {\bf L} = {\bf I} {\bf M} + {\bf P} = {\bf I}{\bf M}_t.
\end{equation} 
Here, ${\bf M}_t = {\bf M + N}$ is the total closure matrix. The Liouville matrix ${\bf L}$ is defined as ${\bf L}_{ij} \equiv (A_{i}| [A_{j}, H])$.
Note that ${\bf I}$ is  Hermitian and positive definite. ${\bf L}$ is Hermitian under the inner product Eq.(4), which guarantees the causality of GF.

The general idea of projective truncation \cite{Mori1,Mori2,Zwanzig1,Roth1,Rowe1,Tserkovnikov1,Lee1,Lee2} is to 
neglect the orthogonal component $\delta B_{i} $ in Eq.(6), i.e., 
\begin{equation}      \label{9}
 \delta B_i  \approx 0.
\end{equation}
Putting Eqs.(5),(6), and (9) into the EOMs of GF, Eqs.(2) and (3), we obtain 
\begin{equation}      \label{10}
G( \vec{A}|\vec{A}^{\dagger} )_{\omega} \approx \left( \omega {\bf 1} - {\bf M}_{t}^{T} \right)^{-1} { \bf I}^{T} .
\end{equation}
Note that the left side and the right side time derivatives of GF, Eqs.(2) and (3), produce the same equation, respecting the time translational invariance of the equilibrium state. Eq.(10) is equivalent to the matrix form of the Mori-Zwanzig equation with the memory function matrix neglected. It becomes exact as the operator basis covers the complete Liouville space.

For given matrices ${\bf M}_{t}$ and ${\bf I}$, the GF matrix in Eq.(10) can be calculated directly by matrix inversion $\left( \omega {\bf 1} - {\bf M}_{t}^{T} \right)^{-1}$, as done in previous analytical studies,\cite{Roth1} or by numerically solving the generalized eigen-value problem of the pair of matrices $({\bf L}, {\bf I})$,
\begin{equation}      \label{11}
   {\bf L U} = {\bf I U \Lambda}.
\end{equation}
Here ${\bf \Lambda} ={\text diag}\{ \lambda_1, \lambda_2, ..., \lambda_n \}$ is a real diagonal matrix. ${\bf U}$ is the generalized eigen vector matrix which diagonalizes ${\bf M}_{t}$, ${\bf U}^{-1} {\bf M}_{t} {\bf U} = {\bf \Lambda}$. It fulfills the generalized orthonormal relation ${\bf U^{\dagger}I U = 1}$. Eq.(10) can be reformulated in terms of  ${\bf U}$ and ${\bf \Lambda}$ as
\begin{equation}      \label{12}
G( \vec{A}|\vec{A}^{\dagger} )_{\omega} \approx ( {\bf IU} )^{\ast} \left( \omega {\bf 1} - {\bf \Lambda}\right)^{-1} ({\bf IU})^{T}.
\end{equation}
The corresponding spectral function reads
\begin{equation}      \label{13}
   \rho(A_{i}|A_{j}^{\dagger})_{\omega} \approx  \sum_{k} (IU)^{\ast}_{ik} (IU)_{jk} \delta(\omega - \lambda_k).
\end{equation}

The Hermitian matrices ${\bf L}$ and ${\bf I}$ contain the average of operators on the state defined by the density matrix $\rho$ in Eq.(4). 
The calculation of them usually relies on additional approximations which cause the arbitrariness. 
In Ref.~\onlinecite{Fan1}, we proposed a practical and systematic method to calculate the matrices ${\bf I}$ and ${\bf L}$. 
The averages of the kind $\langle A_{j}^{\dagger} A_{i} \rangle$ can be obtained from the corresponding GF via the spectral theorem as
\begin{equation}      \label{14}
\langle A_{j}^{\dagger} A_{i} \rangle  =  \sum_{k} \frac{ (IU)^{\ast}_{ik} (IU)_{jk} }{e^{\beta \lambda_{k}} + 1}.
\end{equation}  
For the average of the type $\langle \hat{O}A_{i} \rangle$ ( $\hat{O}$ is an operator outside the basis set $\{A_{k} \}$ ), we use
\begin{equation}      \label{15}
 \langle \hat{O} A_{i} \rangle \approx  \sum_{k} \frac{(IU)_{ik}^{\ast} \left[ U^{T} \langle \{\vec{A}, \hat{O} \} \rangle\right]_{k} }{e^{\beta \lambda_{k}} + 1}.
\end{equation}
$\langle \hat{O}A_{i} \rangle$ can then be calculated self-consistently from ${\bf I}$, ${\bf U}$, and ${\bf \Lambda}$, provided that the averages $\langle \{ A_{i}, \hat{O} \} \rangle$ ($i=1,2,...,n$) are linear combinations of $\{ \langle A_{k}^{\dagger} A_{p} \rangle \}$.\cite{Roth1} 

Usually Eq.(14) and (15) are sufficient to produce ${\bf I}$ but not ${\bf L}$. Therefore, we introduce the following partial projection approximation for ${\bf L}$.
We first classify the basis operators into two groups, $\{A_1, A_2, ..., A_n \} = \{A_1^{(1)}, A_{2}^{(1)}, ..., A_{m}^{(1)} \} \cup \{A_{m+1}^{(2)}, A_{m+2}^{(2)}, ..., A_{n}^{(2)} \}$. The superscripts $(1)$ and $(2)$ denote subset-$1$ and $2$, respectively. Subset-$1$ is composed of basis operators whose commutators with $H$ close automatically. Subset-$2$ contains the rest basis operators. That is, we have $B_{i}^{(1)} = 0$ ($i=1,2,...,m$) and $B_{j}^{(2)} \neq 0$ ($j=m+1, m+2, ..., n$). Associated with this split of basis, the matrices ${\bf I}$, ${\bf M}$, ${\bf P}$, ${\bf N}$, and ${\bf L}$ all become $2 \times 2$ block matrices. In particular, 
\begin{equation}      \label{16}
  \mathbf{P} = \left(
\begin{array} {cc}
    \mathbf{ 0 }   & \mathbf{P}_{12}  \\
    \mathbf{ 0 }   & \mathbf{P}_{22}    
\end{array} \right),
\end{equation}
where ${\bf P}_{12}$ and ${\bf P}_{22}$ are the projection matrices from $\vec{B}^{(2)}$ to $\vec{A}^{(1)}$ and to $\vec{A}^{(2)}$, respectively. 
Similarly, we have 
\begin{equation}      \label{17}
  \mathbf{L} = \left(
\begin{array} {cc}
    \mathbf{ L }_{11}   & \mathbf{L}_{12}  \\
    \mathbf{ L }_{21}   & \mathbf{L}_{22}    
\end{array} \right).
\end{equation}
Employing the Hermiticity of ${\bf L}$, we proposed the following partial projection approximation to ${\bf L}$,
\begin{equation}       \label{18}
\mathbf{L} \approx  \mathbf{L}^{a} = \left(
\begin{array} {cc}
    \mathbf{ (IM) }_{11}  \,\,  & [ \mathbf{ (IM) }_{21}]^{\dagger}  \\
    \mathbf{ (IM) }_{21}  \,\,  & \frac{1}{2}[ \mathbf{ L}^{a}_{22} + (\mathbf{ L}^{a}_{22})^{\dagger} ]
\end{array} \right).
\end{equation}
Here,
\begin{equation}      \label{19}
\mathbf{ L}^{a}_{22} = ({\bf IM})_{22} + {\bf I}_{21} [ {\bf I}_{11} ]^{-1} {\bf P}_{12}
\end{equation}
and we use the exact expression for $ \mathbf{P}_{12}$,
\begin{equation}      \label{20}
 \mathbf{P}_{12} = \left[ \mathbf{ (IM) }_{21} \right]^{\dagger} - \mathbf{ (IM) }_{12}.
\end{equation}
Under this approximation, the input of the calculation are ${\bf M}$ and ${\bf I}$ matrices only. The precision of results is determined only by the selection of basis operators. Below we will show that the precision is improved systematically with enlarged basis size.

\section{Application to Anderson Impurity Model: Formalism}

In this section, we apply the PTA to the Anderson impurity model (AIM). Taking the NRG results as a reference, we compare the results from three successively larger basis sets: the basis at the level of Hubbard-I approximation~\cite{Hubbard1} (HIA basis), at the level of alloy analogy approximation~\cite{Hubbard2} (AAA basis), and at the level of Lacroix approximation~\cite{Lacroix1} (Lacroix basis). These bases form a chain of sets: HIA basis $\subset$ AAA basis $\subset$ Lacroix basis, so that we can speak of {\it enlarging} the basis. Our aim is to study how the results depend on the basis size and to observe the convergence of results to the exact ones in the large basis limit. The Hamiltonian of the AIM that we will study reads
\begin{eqnarray}      \label{21}
\hat{H} &=& \sum_{k\sigma}\left(\epsilon_{k\sigma}-\mu\right) c_{k\sigma}^{\dag}c_{k\sigma} + \sum_{k\sigma}V_{k\sigma}\left(c_{k\sigma}^{\dag}d_{\sigma}+d_{\sigma}^{\dag}c_{k\sigma}\right)\nonumber\\ &+& \sum_{\sigma} \left(\epsilon_{d}-\mu\right)d_{\sigma}^{\dag}d_{\sigma}+ Un_{d\uparrow}n_{d\downarrow}.
\end{eqnarray}
The denotations are standard. For the hybridization function $\Delta_{\sigma}(\omega ) \equiv \sum_{k} V_{k\sigma}^{2} \delta(\omega - \epsilon_{k \sigma})$, we use the Lorentzian type,
\begin{equation}      \label{22}
  \Delta_{\sigma}(\omega )  = \frac{\Delta \omega_{c}^{2}}{ (\omega + \sigma \delta \omega )^{2} + \omega_c^{2}}.
\end{equation}
Here, $\delta \omega$ is the magnetic bias of the bath electrons, introduced to mimic the ferromagnetic leads in quantum dot systems and the situation of magnetic phase in the dynamical mean-field theory.\cite{Vollhardt1,Kotliar1} $\omega_c = 1.0$ is set as the energy unit.
In accordance with $\Delta_{ \bar{\sigma} }(\omega) = \Delta_{\sigma}(-\omega)$, we assume the following constraints in the parameters of $H$,
\begin{eqnarray}      \label{23}
 && \epsilon_{\bar{k} \bar{\sigma}} = - \epsilon_{k \sigma} ;   \nonumber \\
 &&  V_{\bar{k} \bar{\sigma}} = V_{k \sigma}. 
\end{eqnarray}
The particle-hole symmetry of $H$ is realized at the parameter point
$\epsilon_d = - U/2$ and $\mu = 0$.

For HIA and AAA bases, the projection matrix ${\bf P}$ is calculated analytically without using the partial projection approximation. ${\bf N}$ and $G( \vec{A}|\vec{A}^{\dagger} )_{\omega}$ are obtained by analytically solving the linear equations Eqs.(7) and (10). For AAA basis, additional decoupling approximations are used to calculate some of the averages in ${\bf P}$. The particle-hole symmetry is fulfilled automatically in these approximations. The results for Lacroix basis are taken from Ref.~\onlinecite{Fan1}, where we used a particle-hole symmetric form of the partial projection approximation and solved the GF numerically via the generalized eigen-value formalism Eq.(11) on a discretized bath.

\subsection{HIA Basis}

The conventional HIA for AIM is obtained by truncating the EOM at the second order. The involved operators are selected here to form the HIA basis,
\begin{equation}      \label{24}
\left\{ A_{1}= d_{\sigma}, \,\, A_{2k} = c_{k\sigma}, \,\, A_{3} = n_{\bar{\sigma}}d_{\sigma}\right\}.
\end{equation}
In this basis set, $k$ goes through all $n_{k}$ wave vectors of the conduction electron, giving the HIA basis a dimension of $d = 2 + n_{k}$. Due to the conservation of total number of electrons $\hat{N}$ and the $z$ component of total spin $S_{z}$, the basis operators are confined to the type that annihilates an electron with spin $\sigma$. Here we did not use the full spin SU(2) symmetry of AIM. The inner product matrix of the basis operators is
\begin{equation}      \label{25}
{ \bf I}= \left(
\begin{array}{ccc}
1 & 0 & \langle n_{\bar{\sigma}}\rangle\\
0 & \delta_{kp} & 0\\
\langle n_{\bar{\sigma}}\rangle & 0 &\langle n_{\bar{\sigma}}\rangle
\end{array}
\right)
\end{equation}
In Eq.(25), the rank of the matrix corresponds to the sequence $\left\{A_{1}, A_{2k}, A_{3}  \right\}$ and the column corresponds to $\left\{A_{1}, A_{2p}, A_{3}  \right\}$. For simplicity, the full $d \times d$ matrix is abbreviated as a $3 \times 3$ matrix, in which the sub-matrix involving bath operators is abbreviated to a $k$-dependent number. For an example, $\delta_{kp}$ is used to represent the unity sub-matrix $(A_{2k}|A_{2p})$ ($k, p = 1, 2, ..., n_{k}$). Below, our matrix expression will always use this abbreviation convention.

The commutators of the basis operators with $H$ are summarized in Appendix.
The generated new operators are $B_{1}=B_{2k}= 0$ and 
\begin{eqnarray}      \label{26}
&& B_{3} =  \sum_{k}\left[ V_{k\sigma} (n_{\bar{\sigma}}-\frac{1}{2} ) c_{k\sigma}
- V_{k\bar{\sigma}}c_{k\bar{\sigma}}^{\dag}d_{\bar{\sigma}}d_{\sigma} + V_{k\bar{\sigma}}d_{\bar{\sigma}}^{\dag} c_{k\bar{\sigma}} d_{\sigma} \right]. \nonumber \\
&&
\end{eqnarray} 
The matrices ${\bf M}$ and ${\bf P}$ are written as
\begin{eqnarray}      \label{27}
{\bf M} &=&
\left(\begin{array}{ccc}
\epsilon_{d}-\mu & V_{p\sigma} & 0 \\
V_{k\sigma} & (\epsilon_{k\sigma}-\mu)\delta_{kp}  & \frac{1}{2}V_{k \sigma} \\
U  &  0  & \epsilon_{d}-\mu+U
\end{array}
\right),
\end{eqnarray}
and
\begin{equation}      \label{28}
{\bf P} = \left(
\begin{array}{ccc}
0 & 0 & 0\\
0 & 0 & V_{k\sigma} ( \langle n_{\bar{\sigma}}\rangle -1/2 ) \\
0 & 0 & \beta_{\sigma}
\end{array}
\right) , 
\end{equation}
where $\beta_{\sigma} = \sum_{k}V_{k\bar{\sigma}}\langle c_{k\bar{\sigma}}^{\dag}d_{\bar{\sigma}}\left(2n_{\sigma}-1\right)\rangle$. Note that we have included an additional term $-\frac{1}{2}\sum_{k}V_{k \sigma}c_{k \sigma}$ into $B_3$ to make it particle-hole symmetric. The averages are taken as real numbers, {\it i.e.}, $\langle \hat{O}\rangle=\langle \hat{O}^{\dagger} \rangle$.

For this simple case, the full projective truncation can be carried out, {\it i.e.}, ${\bf P}$ (i.e., $\langle n_{\bar{\sigma}} \rangle$ and $\beta_{\sigma}$) can be calculated self-consistently without further approximation. Solving Eqs.(7) and (10) analytically, we obtain the impurity self-energy as
\begin{equation}      \label{29}
\Sigma_{\sigma}(\omega) = U\langle n_{\bar{\sigma}} \rangle + \frac{U^{2} \langle n_{\bar{\sigma}}\rangle \left( 1- \langle n_{\bar{\sigma}}\rangle \right)}{\omega + \mu - \epsilon_{d} - U(1-\langle n_{\bar{\sigma}}\rangle) - \widetilde{\beta}_{\sigma} }.
\end{equation} 
Here $\tilde{\beta}_{\sigma} = \beta_{\sigma} / \left[ \langle n_{\bar{\sigma}}\rangle \left(1-\langle n_{\bar{\sigma}}\rangle \right) \right]$.
Eq.(29) has the form of atomic limit, same as the conventional HIA,\cite{Gebhard1} but with an additional spin-dependent shift $\tilde{\beta}_{\sigma}$ of the impurity level. It is exactly the extended continued fraction expression with the memory function omitted at this level.\cite{Ma1} The impurity GF is obtained from the Dyson equation $G(d_{\sigma}|d_{\sigma}^{\dagger})_{\omega} = \left[  G_{0 \sigma}^{-1}(\omega) - \Sigma_{\sigma}(\omega) \right]^{-1}$. We also obtain $G(n_{\bar{\sigma}}d_{\sigma}|d_{\sigma}^{\dagger})_{\omega} = (1/U) G(d_{\sigma}|d_{\sigma}^{\dagger})_{\omega} \Sigma_{\sigma}(\omega)$.  The non-interacting impurity GF is given by
\begin{equation}      \label{30}
G_{0 \sigma}(\omega) = \frac{1}{\omega + \mu - \epsilon_{d} - \Gamma_{\sigma}(\omega + \mu) },
\end{equation}
with $\Gamma_{\sigma}(\omega) = \int_{-\infty}^{+\infty} \Delta_{\sigma}(\epsilon) / (\omega - \epsilon) d\epsilon$.
For the averages, $\langle n_{\bar{\sigma}}\rangle$ can be calculated from $G(d_{\bar{\sigma}}|d_{\bar{\sigma}}^{\dagger})_{\omega}$. $\tilde{\beta}_{\sigma}$ needs to be calculated from the right-hand side EOMs
\begin{eqnarray}      \label{31}
&&  G( d_{\bar{\sigma}} | c_{k\bar{\sigma}}^{\dag} )_{\omega} = \frac{V_{k\bar{\sigma}}}{\omega+\mu-\epsilon_{k\bar{\sigma}}} G( d_{\sigma}|d_{\sigma}^{\dagger})_{\omega},      \nonumber \\
&&  G( n_{\sigma}d_{\bar{\sigma}} | c_{k\bar{\sigma}}^{\dag} )_{\omega} = \frac{V_{k\bar{\sigma}}}{\omega+\mu-\epsilon_{k\bar{\sigma}}} G(n_{\bar{\sigma}}d_{\sigma}|d_{\sigma}^{\dagger})_{\omega}.
\end{eqnarray}
For a paramagnetic bath and at the particle-hole symmetric point, $\tilde{\beta}_{\sigma} =0$. Eq.(29) recovers that of the conventional HIA. Away from the particle-hole symmetry or for $\delta \omega \neq 0$, $\tilde{\beta}_{\sigma} \neq 0$ and Eq.(29) differs from the conventional HIA.

\subsection{AAA basis}

The AAA basis is obtained by adding the operators $A_{4k} = n_{\bar{\sigma}} c_{k\sigma}$ ($k=1,2,..., n_{k}$) into the HIA basis,
\begin{equation}      \label{32}
\left\{ A_{1}= d_{\sigma}, \,\, A_{2k} = c_{k\sigma}, \,\,  A_{3} = n_{\bar{\sigma}}d_{\sigma}, \,\,  A_{4k} = n_{\bar{\sigma}} c_{k\sigma} \right\}.
\end{equation}
The total dimension is $d=2+2n_{k}$. The Tyablikov-type decoupling of EOM at this level results in an approximation which, when combined with dynamical mean-field theory for Hubbard model, gives the conventional AAA.\cite{Hubbard2,Gebhard1} Similar calculation is carried out as for the HIA basis. The inner product matrix reads
\begin{equation}      \label{33}
{ \bf I}= \left(
\begin{array}{cccc}
1 &  0  & \langle n_{\bar{\sigma}}\rangle  &  0\\
0 & \delta_{kp} & 0 & \langle n_{\bar{\sigma}} \rangle\delta_{kp}\\
\langle n_{\bar{\sigma}}\rangle & 0 &\langle n_{\bar{\sigma}}\rangle & 0 \\
0 & \langle n_{\bar{\sigma}} \rangle\delta_{kp} & 0 & \langle n_{\bar{\sigma}}\rangle \delta_{kp}
\end{array}
\right) .
\end{equation}

Using the commutators summarized in Appendix, we obtain the matrix ${\bf M}$ as 
\begin{eqnarray}      \label{34}
{\bf M} &=&
\left(\begin{array}{cccc}
\epsilon_{d}-\mu & V_{p\sigma} & 0  &  0 \\
V_{k\sigma} & (\epsilon_{k\sigma}-\mu)\delta_{kp}  & 0  & 0 \\
U  &  0  & \epsilon_{d}-\mu+U  & V_{p \sigma}  \\
0  &  0  & V_{k \sigma}  & (\epsilon_{k\sigma}-\mu)\delta_{kp}
\end{array}
\right) .  \nonumber \\
&&
\end{eqnarray}
Commutators of $A_{3}$ and $A_{4k}$ with $H$ generate the new operators $B^{(2)}_{3}$ and $B^{(2)}_{4k}$,
\begin{eqnarray}      \label{35}
B^{(2)}_{3} &=& \sum_{k}V_{k\bar{\sigma}} ( d_{\bar{\sigma}}^{\dag} c_{k\bar{\sigma}} d_{\sigma} - c_{k\bar{\sigma}}^{\dag}d_{\bar{\sigma}}d_{\sigma} ),   \nonumber \\
B^{(2)}_{4k} &=& \sum_{p} V_{p\bar{\sigma}} ( d_{\bar{\sigma}}^{\dag} c_{p\bar{\sigma}} c_{k\sigma} - c_{p\bar{\sigma}}^{\dag}d_{\bar{\sigma}}c_{k\sigma}  ).
\end{eqnarray} 
The projection matrix ${\bf P}$ in the block form of Eq.(16) has sub-matrices ${\bf P}_{11} ={\bf P}_{21}={\bf P}_{12}=0$, and 
\begin{eqnarray}      \label{36}
 ({\bf P}_{22})_{3,3} &=& \sum_{k}V_{k\bar{\sigma}} \langle c_{k \bar{\sigma}}^{\dag}d_{\bar{\sigma}}(2n_{\sigma}-1)\rangle  , \nonumber \\
 ({\bf P}_{22})_{3, 4p} &=&   ({\bf P}_{22})_{4p, 3}  \nonumber \\
   &=& \sum_{k} V_{k\bar{\sigma}} \langle (c_{k\bar{\sigma}}^{\dag}d_{\bar{\sigma}} + d_{\bar{\sigma}}^{\dag} c_{k\bar{\sigma}} ) c_{p\sigma}^{\dag}d_{\sigma}  \rangle    ,   \nonumber \\
 ({\bf P}_{22})_{4k,4p} &=& \sum_{q} V_{q\bar{\sigma}} \langle ( c_{k \sigma}^{\dag}c_{p\sigma} - \frac{1}{2} \delta_{kp} ) ( c_{q \bar{\sigma}}^{\dag} d_{\bar{\sigma}} + d_{\bar{\sigma}}^{\dag} c_{q\bar{\sigma}}) \rangle .  \nonumber \\
 &&
\end{eqnarray}

For the AAA basis, it is still possible to analytically solve the linear equations Eqs.(7) and (10) for the local GFs $G(d_{\sigma}|d_{\sigma}^{\dagger})_{\omega}$ and $G(n_{\bar{\sigma}}d_{\sigma}|d_{\sigma}^{\dagger})_{\omega}$. However, the averages in $({\bf P}_{22})_{3, 4p}$ and $({\bf P}_{22})_{4k, 4p}$ cannot be written in the form $\langle A_{i}^{\dagger} A_{j} \rangle$. If we use the partial projection approximation Eq.(18) for ${\bf L}$,  due to ${\bf P}_{12}=0$, we recover the conventional AAA which is equivalent to setting $B^{(2)}_{3} = B^{(2)}_{4k} \approx 0$. To go beyond the conventional AAA, we use a simple decoupling approximation for the elements of ${\bf P}_{22}$,
\begin{eqnarray}      \label{37}
 ({\bf P}_{22})_{3, 4p} &=&   ({\bf P}_{22})_{4p, 3} \approx 2 \sum_{k} V_{k\bar{\sigma}} \langle c_{k\bar{\sigma}}^{\dag}d_{\bar{\sigma}} \rangle \langle c_{p\sigma}^{\dag}d_{\sigma}  \rangle ,   \nonumber \\
 ({\bf P}_{22})_{4k,4p}  & \approx & 0 .
\end{eqnarray}
Since this approximation keeps ${\bf P}_{22}$ symmetric and ${\bf IM}$ is also symmetric, the Hermiticity of ${\bf L}$ is conserved. The particle-hole symmetry is also fulfilled.

Solving Eqs.(7) and (10) analytically, we obtain the self-energy for AAA basis as
\begin{eqnarray}      \label{38}
&& \Sigma_{\sigma}(\omega) \nonumber \\
&=& U \langle n_{\bar{\sigma}} \rangle +  \frac{ U^2 \langle n_{\bar{\sigma}}\rangle \left( 1- \langle n_{\bar{\sigma}}\rangle \right)}{\omega + \mu - \epsilon_{d} - U(1 - \langle n_{\bar{\sigma}}\rangle) - \tilde{\beta}_{\sigma} - \mathcal{B}_{\sigma}(\omega)  } . \nonumber \\
&&
\end{eqnarray}
Compared to the self-energy of HIA basis, a frequency-dependent shift and broadening of the impurity level appears as
\begin{equation}      \label{39}
\mathcal{B}_{\sigma}(\omega) = \sum_{p} \frac{\left(V_{p \sigma} + B_{p \sigma}\right)^{2}}{\omega +\mu - \epsilon_{p \sigma}},
\end{equation}
with 
\begin{equation}      \label{40}
 B_{p \sigma} = \frac{ 2 \left[ \sum_{k} V_{k \bar{\sigma}} \langle c_{k\bar{\sigma}}^{\dagger} d_{\bar{\sigma}} \rangle \right] \langle c_{p\sigma}^{\dag} d_{\sigma}\rangle }{ \langle n_{\bar{\sigma}}\rangle \left(1 - \langle n_{\bar{\sigma} } \rangle  \right) }. 
\end{equation}
Neglecting $B_{p\sigma}$ and $\tilde{\beta}_{\sigma}$, Eq.(38) recovers the conventional AAA.\cite{Gebhard1} Eq.(38) is also consistent with the form of extended continued fraction,\cite{Ma1} but with an approximate expression for the memory function.
Using EOM of GF $G(d_{\sigma}|c_{p\sigma}^{\dagger } )_{\omega}$, we reduce Eq.(39) to
\begin{eqnarray}      \label{41}
&&   \mathcal{B}_{\sigma}(\omega) =  \int_{-\infty}^{+\infty} d\epsilon \frac{\Delta_{\sigma}(\epsilon)}{\omega + \mu - \epsilon}  \times   \nonumber \\
&&  \left[1 + \frac{2 \langle \Gamma_{\bar{\sigma}}(\omega + \mu) G_{\bar{\sigma}}(\omega) \rangle}{ \langle n_{\bar{\sigma}}\rangle - \langle n_{\bar{\sigma} }\rangle^{2}} \Big\langle \frac{G_{\sigma}(\omega)}{\omega + \mu -\epsilon} \Big\rangle \right]^{2}.  
\end{eqnarray} 
Here, $G_{\sigma}(\omega) = G(d_{\sigma}| d_{\sigma}^{\dagger} )_{\omega}$. The symbol $\langle g(\omega) \rangle$ represents $ (-1/\pi) \int_{-\infty}^{+\infty} {\text Im} [ g(\omega+ i \eta)] / ( e^{\beta \omega} + 1 ) d\omega $ with $\eta$ being an infinitesimal positive number.
In Eq.(39), the shift of $V_{k \sigma}$ is generated by projecting $B_{3}$ to $A_{4k}$ and $B_{4k}$ to $A_{3}$. It contains the spin exchange between impurity and bath electrons. As will be shown below, this renormalization of hybridization produces improved description of the Kondo peak at low temperatures compared to conventional AAA.

\subsection{Lacroix basis}

In the work of Lacroix,\cite{Lacroix1} the GFs generated by the commutator of $H$ and $n_{\bar{\sigma}} d_{\sigma}$ are kept and the truncation is done in the next order EOM. Here we take the corresponding operators to form the Lacroix basis $\left\{ A_{1}, A_{2k}, A_{3}, A_{4k}, A_{5k}, A_{6k} \right\}$ ($k=1,2,..., n_{k}$), with
\begin{eqnarray}      \label{42}
&&  A^{(1)}_{1}= d_{\sigma}, \,\,\, A^{(1)}_{2k} = c_{k\sigma}, \,\,\, A^{(1)}_{3} = n_{\bar{\sigma}}d_{\sigma},   \nonumber \\
&& A^{(2)}_{4k} = n_{\bar{\sigma}} c_{k\sigma}, \,\,\, A^{(2)}_{5k} = d_{\bar{\sigma}}^{\dagger} c_{k \bar{\sigma}} d_{\sigma}, \,\,\, A^{(2)}_{6k} = c_{k\bar{\sigma}}^{\dagger} d_{\bar{\sigma}} d_{\sigma} . \nonumber \\
&&
\end{eqnarray}
The superscripts $(1)$ and $(2)$ denote the grouping of basis operators according to the closure properties of their commutators with $H$: $B^{(1)}_{i} = 0$ and $B^{(2)}_{i} \neq 0$. For this basis, we directly take the results from Ref.~\onlinecite{Fan1}, where we used the particle-hole symmetric partial projection truncation. We numerically solved the PTA equations for a linearly-discretized bath with $n_{k}=401$ bath sites, which already represents the continuous bath satisfactorily. The half band width is $D=5.0$. The $\delta$-peaks in the LDOS were broadened with $\eta=0.01 \sim 0.02$. 

\section{Numerical Results and Comparison}

Using the formalism in previous sections, we obtain numerical results for HIA, AAA, as well as Lacroix bases. Below, these approximations are called projective-HIA (pHIA), projective-AAA (pAAA), and projective-Lacroix (pLacroix), respectively. The NRG results, used as a reference, are obtained from the full density matrix algorithm\cite{Weichselbaum1,Peters1}. For the local density of states (LDOS), we use the self-energy trick~\cite{Bulla2} and average on $N_{z}=8$ interleaved discretizations.~\cite{Yoshida1} The logarithmic discretization parameter is $\Lambda=2.0$ and we keep $M_s = 350 \sim 380$ states.  Though not extrapolated to the exact limit $\Lambda=1$ and $M_{s}=\infty$,~\cite{Li1} we have checked that the uncertainties in NRG results are much smaller than the difference between NRG and all the approximate results. For the results below, we fix $\mu=0.0$ and $\Delta=0.1$.

\begin{figure}[t!]
\vspace{-1.0cm}
\begin{center}
\includegraphics[width=5.1in, height=3.9in, angle=0]{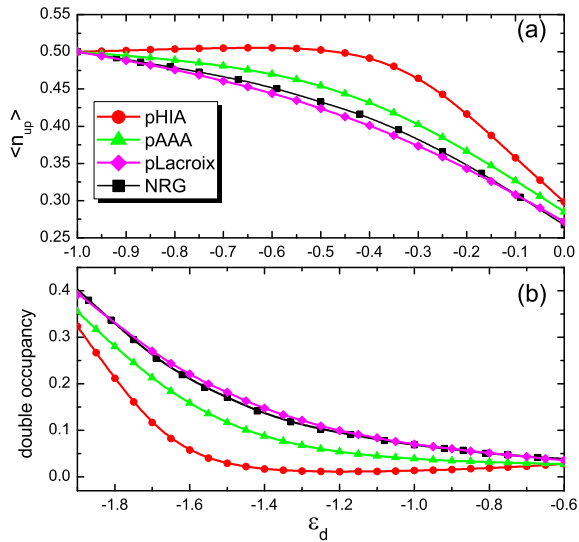}
\vspace*{-2.5cm}
\end{center}
\caption{ (a) $\langle n_{\uparrow} \rangle$ and (b) $\langle n_{\uparrow} n_{\downarrow}\rangle$ as functions of $\epsilon_d$. The parameters are $U=2.0$, $T=0.1$, $\delta \omega = 0.0$.  }   \label{Fig1}
\end{figure}

We first study the impurity electron occupation $\langle n_{\sigma} \rangle$ and the double occupancy $\langle n_{\uparrow} n_{\downarrow} \rangle$ as functions of $\epsilon_d$, $\delta \omega$, $U$, and $T$. They describe the static magnetic and the charge response of the impurity to external parameters.
In Fig.1, we plot $\langle n_{\uparrow} \rangle$ (Fig.1(a)) and $\langle n_{\uparrow} n_{\downarrow} \rangle$ (Fig.1(b)) as functions of $\epsilon_d$, for a non-magnetic bath at $U=2.0$ and $T=0.1$. The curves of pHIA, pAAA, and pLacroix are compared with those of NRG. 
As the basis is enlarged from HIA to Lacroix, both quantities shift towards NRG results in the whole $\epsilon_d$ regime, with slight overshooting in the pLacroix results. At $\epsilon_d=-1.0$, all methods give $\langle n_{\uparrow}\rangle=0.5$ due to the particle-hole symmetry. The significant improvement of pLacroix over pHIA and pAAA shows that the hybridization effect lacking in the HIA and AAA bases is important for quantitative accuracy.

\begin{figure}[t!]
\vspace{-1.0cm}
\begin{center}
\includegraphics[width=5.1in, height=3.9in, angle=0]{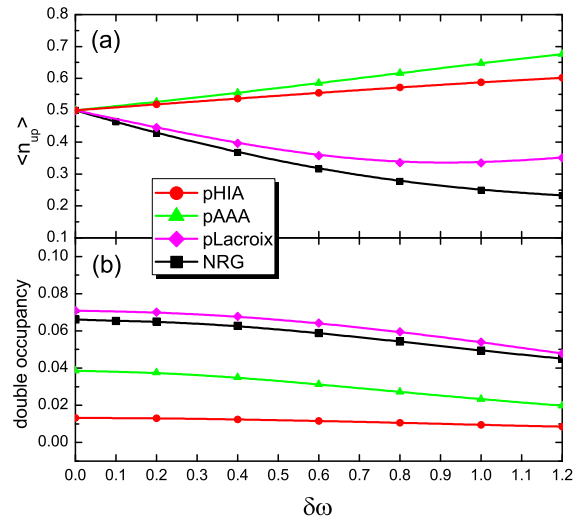}
\vspace*{-2.5cm}
\end{center}
\caption{(a) $\langle n_{\uparrow} \rangle$ and (b) $\langle n_{\uparrow} n_{\downarrow}\rangle$ as functions of $\delta \omega$. The parameters are $U=2.0$, $T=0.1$, $\epsilon_d=-U/2$. }  
\label{Fig2}
\end{figure}

In Fig.2, we plot $\langle n_{\uparrow} \rangle$ and $\langle n_{\uparrow} n_{\downarrow} \rangle$ as functions of $\delta \omega$ for $U=2.0$, $T=0.1$, and $\epsilon_d = -U/2$. The particle-hole symmetry at this parameter set assures the exact $\langle n_\uparrow \rangle = 0.5$ at $\delta \omega=0$. Away from $\delta \omega=0$, the deviation $\langle n_\uparrow \rangle$ begins to increase for all bases, but pLacroix gives the smallest deviation. In particular, pLacroix gives the correct sign in the impurity spin response to the bath bias. The double occupancy shown in Fig.2(b) has a weak $\delta \omega$ dependency. Again, we observe that the results from projective truncations tends to those of NRG systematically with increasing basis size.

\begin{figure}[t!]
\vspace{-1.0cm}
\begin{center}
\includegraphics[width=5.2in, height=3.9in, angle=0]{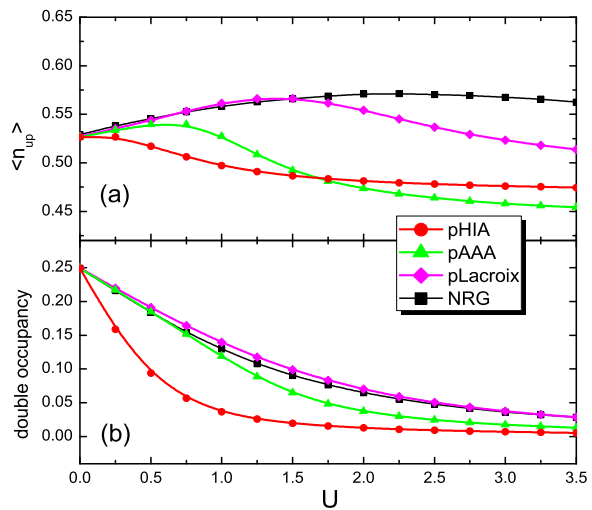}
\vspace*{-2.5cm}
\end{center}
\caption{ (a) $\langle n_{\uparrow} \rangle$ and (b) $\langle n_{\uparrow} n_{\downarrow}\rangle$ as functions of $U$. The parameters are $T=0.1$, $\delta \omega = -0.2$, and $\epsilon_d = -U/2$. }   \label{Fig3}
\end{figure}
Fig.3 shows the same averages as functions of $U$ for $T=0.1$, $\epsilon_d = -U/2$, and a negative bath bias $\delta\omega = -0.2$. At $U=0.0$, all projection truncations give exact results for $\langle n_{\uparrow} \rangle$ and $\langle n_{\uparrow}n_{\downarrow} \rangle$. In Fig.3(a), as $U$ increases from zero, the agreement with the NRG curve is maintained to larger $U$ values for larger basis, up to $U=1.5$ for pLacroix. In Fig.3(b), pHIA gives significant deviation in $\langle n_{\uparrow}n_{\downarrow} \rangle$ as soon as $U > 0$. pAAA gives good agreement up to $U=1.0$. The result from pLacroix is not as accurate as pAAA in the small $U$ regime but the overall agreement, especially in the intermediate to large $U$ regime, is much better.

\begin{figure}[t!]
\vspace{-1.0cm}
\begin{center}
\includegraphics[width=5.3in, height=4.0in, angle=0]{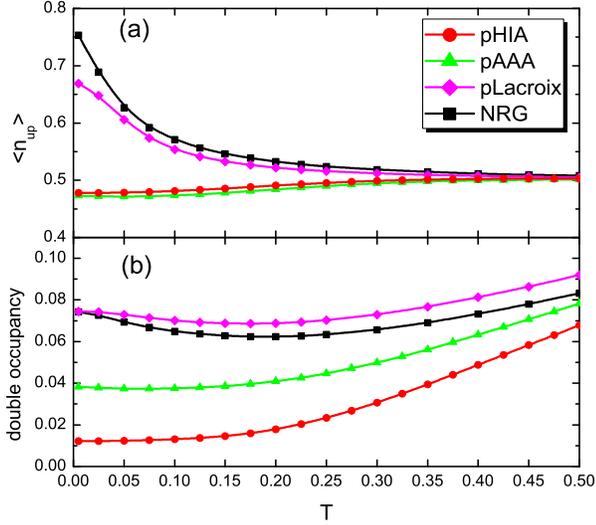}
\vspace*{-3.0cm}
\end{center}
\caption{(a) $\langle n_{\uparrow} \rangle$ and (b) $\langle n_{\uparrow} n_{\downarrow}\rangle$ as functions of temperature.  The parameters are $U=2.0$, $\delta \omega = -0.2$, $\epsilon_d=-U/2$. }   \label{Fig4}
\end{figure}

\begin{figure}[t!]
\vspace{-1.0cm}
\begin{center}
\includegraphics[width=4.8in, height=3.8in, angle=0]{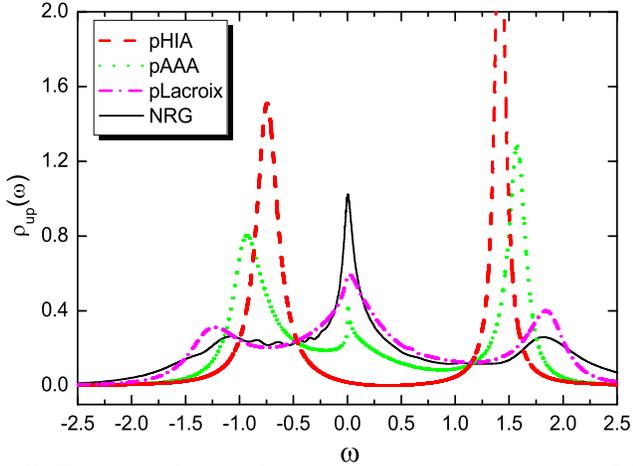}
\vspace*{-3.0cm}
\end{center}
\caption{ Impurity density of states calculated at $U=2.0$, $T=0.001$, $\delta \omega =0.0$, $\epsilon_d = -0.7$.   }   \label{Fig5}
\end{figure}

The temperature dependence of the same quantities are shown in Fig.4 for $U=2.0$, $\epsilon_d=-U/2$, $\delta\omega = -0.2$. In Fig.4(a), using the particle-hole symmetry properties $\langle n_{\uparrow}\rangle + \langle n_{\downarrow} \rangle=1$, we can deduce that the impurity spin polarization is zero at $T=\infty$ and it increases as temperature is lowered for all approximations. While pHIA and pAAA give $\langle n_{\uparrow} \rangle < 0.5$ which has the wrong sign of spin polarization, pLacroix gives correct sign and quantitative agreement with NRG for all temperatures. In Fig.4(b), $\langle n_{\uparrow} n_{\downarrow}\rangle$ decreases with decreasing temperature but a slight increase is observed in both pAAA and pLacroix curves (below $T=0.05$ for pAAA and $T=0.15$ for pLacroix). This upturn of double occupancy, also seen in NRG result, reflects the screening of local moment and forming of the Fermi liquid state below the Kondo temperature. It is notable that the double occupancy from pLacroix is very accurate at $T=0$. Similar pattern of convergence is observed in these data as the basis is enlarged from HIA to Lacroix.

Fig.5 shows the LDOS obtained from different bases at a particle-hole asymmetric point $\epsilon_d=-0.7$, $U=2.0$, and at low temperature $T=0.001$. They are compared with NRG result. The LDOS from various PTAs correctly contain the upper and lower Hubbard peaks. As the basis is enlarged from HIA to Lacroix, both the position and the weight of the Hubbard peaks tend to those of NRG systematically. At $\omega = 0$, pHIA does not produce a Kondo peak while pAAA and pLacroix produce a sharp Kondo peak. Quantitatively comparing the weight and the shape of Kondo peaks, pAAA gives a too much sharper peak with small weight while pLacroix produces a slightly broader peak with closer weight to NRG. The overall tendency of the convergence in LDOS is apparent.

\begin{figure}[t!]
\vspace{-3.5cm}
\begin{center}
\includegraphics[width=6.0in, height=4.3in, angle=0]{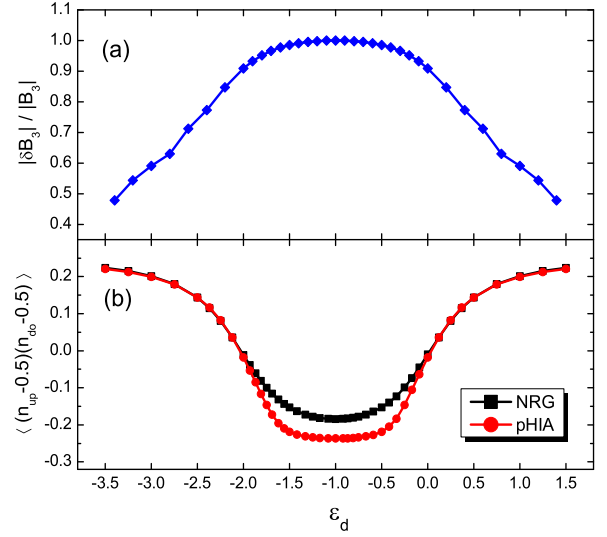}
\vspace*{-1.0cm}
\end{center}
\caption{ (a) relative truncation error $e_{T}$ in pHIA, (b) symmetrized double occupancy, as functions of $\epsilon_d$. Parameters are  $U=2.0$, $T=0.1$, and $\delta \omega =0.0$.   }   \label{Fig6}
\end{figure}

In PTA, using the inner product, we can quantify the truncation error. Here we take the HIA basis as an example, in which the only truncation approximation is $\delta B_{3} \approx 0$. Extension to more complicated truncations is possible. For pHIA, we propose the following quantity to measure the truncation error,
\begin{equation}
   e_{T} = \frac{|\delta B_3|}{|B_3|},
\end{equation}
where $|X| \equiv \left(X|X \right)^{1/2}$ is the norm of operator $X$ under the inner product Eq.(4). $B_3$ and $\delta B_3$ are given by Eq.(26) and Eq.(6), respectively. The Pythagorean theorem implies that $0 \leq e_{T} \leq 1$. At finite temperature, $e_{T}=0$ is equivalent to $\delta B_3 = 0$. In this case no approximation is made and PTA becomes exact. In the other limit, $e_{T}=1$ means $B_{3}\approx 0$, a complete negligence of the new operators produced by EOM.
Therefore, $e_{T}$ is a quantitative gauge of the truncation error in PTA. In Fig.6(a), we plot $e_{T}$ as a function of $\epsilon_d$ for $U=2.0$, $T=0.1$, and $\delta \omega = 0.0$. We use pLacroix to calculate all the inner product involved in Eq.(43), including the projecting matrix ${\bf N}$ of Eq.(6). For comparison, in Fig.6(b), we show the particle-hole symmetrized double occupancies from NRG and pHIA as functions of $\epsilon_d$. 

It is seen from Fig.6 that $e_T$ qualitatively reflects the errors in the physical quantities. At the particle-hole symmetric point $\epsilon_d = -1.0$, $e_{T}=1.0$, being consistent with the fact that at this point, pHIA recovers the conventional HIA which amounts to $B_3 \approx 0$. In the regime $-2.0 \leq \epsilon_d \leq 0.0$, $e_T$ stays close to $1.0$, showing that pHIA has the largest truncation error in this regime. Correspondingly, the discrepancy in the symmetrized double occupancy between NRG and pHIA is significant in this regime. Further away from this regime, $e_{T}$ quickly decreases and in Fig.6(b), the double occupancies from pHIA and NRG merge. In the limits $\epsilon_d = \pm \infty$, $\langle n_{\sigma}\rangle = 0$ or $1$, we expect $e_{T}=0$. pHIA will become exact since no electron correlation is present in those limits. Fig.6 shows a qualitative correlation between $e_{T}$ and the error in the double occupancy of pHIA. We observe similar correlations in other quantities as well, but there is no strict monotonous correspondence between $e_{T}$ and the errors. This is because in PTA, the physical quantities depend on the truncation error non-linearly. Therefore, we conclude that $e_{T}$ can be used to gauge the overall level of approximation of PTA. It is especially suitable for comparing the error among different parameter regimes.

\section{Discussion and Summary}

First, let us discuss the scaling of error with the basis size. We did not study this scaling quantitatively because for AIM with a continuous bath, it is difficult to quantify the size of basis. In fact, for pHIA and pAAA, the dimension of the basis is infinity if we count the number of linearly independent operators in the basis, due to infinitely many bath modes in AIM. Therefore, below we only give a qualitative discussion of this issue. If we regard truncating the EOM of GFs as such a problem: for an original operator $A$ in the full operator space $\mathcal{S}$, look for the operator $A^{\prime}$ in a subspace $\mathcal{S}^{\prime} \subset \mathcal{S}$ and require that $A^{\prime}$ is as close to $A$ as possible. The solution $A^{\prime}$ will be the projection of $A$ into $\mathcal{S}^{\prime}$. In this sense, for a given inner product, projective truncation is the optimal way of truncating the chain of EOMs. This is why the PTA could be superior to conventional decoupling in accuracy. However, in PTA, the inner product $\left( X|Y \right)$ is not calculate exactly (except for those trivial ones such as $(d_{\sigma}|d_{\sigma})$) but self-consistently from the GFs obtained in this theory. As a result, even without partial projection approximation, PTA has two sources of error, Liouville space truncation and the approximate evaluation of projection. As the basis is enlarged, both the space truncation and the precision of projection are improved. Thus we expect that the accuracy in the final result improves beyond linear fashion with the basis size. 

With enlarging basis, the rate of convergence in results depends crucially on the basis selection method. Here, we simply collect those separate operators appearing in the successive EOMs, $\left[A_{1}, H\right]$, $\left[ \left[A_{1}, H\right], H \right]$, etc. There are other ways of selecting basis operators.~\cite{Avella3} Especially, the selection of orthogonal basis operators in the Krylov subspace produces a continued fraction form for the local GF.~\cite{Lee1,Lee2} The self-energy functions from pHIA and pAAA in this work have the extended continued fraction form~\cite{Ma1} used to do resummation for the strong-coupling series expansions of GF.~\cite{Tong1} The efficiency of basis, measured by the accuracy versus basis size scaling, depends on how fast the key excitations are taken into account as the basis is enlarged. Finding the optimal basis selection procedure is an important research topic for the future.

Up to now, the best results that we obtain for AIM is from pLacroix. From Fig.2 and Fig.3, it is clear that even for pLacroix, the accuracy in the impurity spin polarization under the bath bias is not satisfactory, especially in the large $U$ and large $\delta \omega$ regime which is important for describing the antiferromagnetic phase in Hubbard model within DMFT. Therefore, it is natural to go beyond Lacroix basis. However, we find that it is not easy to maintain the positive definiteness of the inner product matrix ${\bf I}$ for larger basis set, possibly due to the inclusion of basis operators which has very small norm. Method such as singular value decomposition is being considered to removed those excitation modes of tiny norm.

In summary, in this paper we compare the PTA results obtained from three successively larger bases with those from NRG. The results improve systematically with increasing basis size and a clear tendency of convergence to NRG results is observed. We also propose a quantity to gauge the truncation error in PTA and demonstrate its usefulness in pHIA. Our results confirm that the PTA is a computational method of controllable precision for quantum many-body systems.

\section{Acknowledgements}
This work is supported by 973 Program of China (2012CB921704), NSFC grant (11374362), Fundamental Research Funds for the Central Universities, and the Research Funds of Renmin University of China 15XNLQ03.

\appendix{}

\section{Commutators of $[A_{i}, H]$ }

In this Appendix, we summarize the commutators between the basis operators and the AIM Hamiltonian $H$ Eq.(21). Below, we give the commutators for HIA and AAA bases. For Lacroix basis, see Appendix B of Ref.~\onlinecite{Fan1}.
\begin{eqnarray}      \label{B1-B4}
 \left[ d_{\sigma}, H \right] &=& (\epsilon_d - \mu) d_{\sigma} + \sum_{k} V_{k\sigma} c_{k\sigma} + U n_{\bar{\sigma}} d_{\sigma},   \\
\left[ c_{k \sigma}, H \right] &=& (\epsilon_{k \sigma} - \mu) c_{k\sigma} + V_{k \sigma} d_{\sigma},   \\
\left[ n_{\bar{\sigma}} d_{\sigma}, H \right]  &=&  (\epsilon_{d} - \mu +U ) n_{\bar{\sigma}} d_{\sigma} + \sum_{k} V_{k\sigma} n_{\bar{\sigma}} c_{k\sigma}  \nonumber \\
     && + \sum_{k} V_{k \bar{\sigma}} ( d_{\bar{\sigma}}^{\dag} c_{k\bar{\sigma}} - c_{k\bar{\sigma}}^{\dagger} d_{\bar{\sigma}} ) d_{\sigma},  \\
\left[ n_{\bar{\sigma}} c_{k \sigma}, H \right] &=& (\epsilon_{k \sigma} - \mu) n_{\bar{\sigma}} c_{k \sigma} + V_{k \sigma} n_{\bar{\sigma}} d_{\sigma} \nonumber \\
 && + \sum_{p} V_{p \bar{\sigma}} ( d_{\bar{\sigma}}^{\dag} c_{ p \bar{\sigma}} - c_{p \bar{\sigma}}^{\dagger} d_{\bar{\sigma}} ) c_{k \sigma}.  
\end{eqnarray}

\end{document}